\documentclass[10pt,pra,noshowpacs,twocolumn,floatfix,superscriptaddress]{revtex4}  

\usepackage{graphics}
\usepackage{epsfig}

\begin{document}

\title{Tuning the Thermal Expansion Properties of Optical Reference Cavities with Fused Silica Mirrors}

\author{Thomas Legero}
\email[Corresponding author ]{Thomas.Legero@ptb.de}
\author{Thomas Kessler}
\author{Uwe Sterr}
\affiliation{Physikalisch--Technische Bundesanstalt (PTB), Bundesallee 100, 38116 Braunschweig, Germany}

\begin{abstract}

We investigate the thermal expansion of low thermal noise {F}abry-{P}\'erot cavities made of Low Thermal Expansion (LTE) glass spacers and Fused Silica (FS) mirrors. The different thermal expansion of mirror and spacer deforms the mirror. This deformation strongly contributes to the cavity's effective Coefficient of Thermal Expansion (CTE), decreasing the zero crossing temperature by about 20 K compared to an all-LTE glass cavity. Finite element simulations and CTE measurements show that LTE rings optically contacted to the back surface of the FS mirrors allow to tune the zero crossing temperature over a range of 30 K.    

\end{abstract}

\maketitle 

\section{Introduction}

Ultrastable optical cavities have become a standard tool for stabilizing laser systems needed for high-resolution spectroscopy, optical clocks \cite{ros08,lud08}, optical microwave generation \cite{bar05,mil09a, lip09} and coherent optical frequency transfer \cite{wil08a,jia08}. State-of-the-art cavity-stabilized laser systems show linewidths below 1 Hz and fractional frequency instabilities on the order of $10^{-15}$ at one second \cite{you99,sto06,not06,lud07,web08,dub09}. 

The fractional length stability of typical 10 cm long cavities is limited to a flicker floor at the $10^{-15}$ level which is attributed to inevitable thermal noise (Brownian fluctuation) of the cavity \cite{num04}. The power spectral density of these length fluctuations $S_L$ is proportional to the cavity temperature $T$ and to the mechanical loss factors of the cavity materials. The length fluctuations translate into frequency fluctuations by $S_\nu = S_L \, \nu^2/L^2$, with the laser frequency $\nu$ and the cavity length $L$. 

There are several options to decrease the cavity's thermal noise. A first way is cooling the cavity to cryogenic temperatures. In this case the cavities are made of sapphire or silicon, since at low temperatures both materials show a zero crossing in their CTE and the stabilization of the cavity temperature at this zero-CTE point greatly improves the thermal length stability. However, handling cryogenic temperatures demands  a bigger, more complex and more expensive set-up than a simple temperature stabilization around $20\,^{\circ}\mathrm{C}$. Especially for transportable systems cryogenic cavities are not appropriate.

For room temperature applications different glasses and glass-ceramics are available which show a zero crossing of their CTE around $20\,^{\circ}\mathrm{C}$. Such low thermal expansion materials are e.g. Zerodur, Corning ULE, Asahi AZR and Clear Ceram \cite{tradename}. The thermal noise of such cavities can be reduced by significantly increasing the cavity length $L$, but longer cavities are more sensitive to accelerations $a_\mathrm{vib}$ from mechanical vibrations. Since the short-term length stability is mainly affected by seismic and acoustic noise there has been much effort to reduce the vibration sensitivity. Optimized cavity mounting geometries for cavities of around 10 cm length \cite{not05,naz06} with sensitivities smaller than $(\Delta L / L)/a_\mathrm{vib} = 2 \times 10^{-11}$/(m s$^{-2}$) \cite{web07,mil09} have been realized. 

A different approach is based on the use of low thermal noise materials. Here we focus on fused silica which has a very small mechanical loss at room temperature. A big drawback of this material is its large room temperature CTE of about $500 \times 10^{-9}$/K \cite{ber76}. Therefore, it is not an appropriate choice for the spacer material. However, the most dominant thermal noise of a cavity made of e.g. Cornings ULE glass arises from the mirror substrates (84 \% of $S_\mathrm{L}$ \cite{num04}). Replacing the ULE mirrors of a 10 cm long ULE cavity by FS mirrors can reduce the fractional instability from thermal noise roughly by a factor of three to a flicker floor of a few times $10^{-16}$ \cite{mil09}. 

The CTE mismatch between the LTE spacer and the FS mirrors is a problem since it leads to an axial mirror bending. As a result the effective CTE of the combined material cavity can be quite different from the CTE of the spacer material. This effect has been thoroughly studied for cryogenic cavities made of a sapphire spacer and FS mirrors \cite{not95,won97}. It has also been discussed for ULE cavities where spacer and mirrors are from different batches and may therefore show slightly different coefficients of thermal expansion \cite{fox08}. 

In this paper we investigate the thermal expansion of LTE glass cavities with FS mirrors by means of Cornings ULE as an example of an low thermal expansion material. We show that rings made of ULE glass optically contacted to the back surface of the mirrors compensate the thermal mirror bending. In section \ref{CombMatCav} we analyze the effective CTE of combined material cavities. Finite element simulations reveal a zero crossing temperature shift of the effective cavity CTE of a few 10 K in respect to the zero crossing temperature of an all-ULE cavity. The effect of additional ULE rings is discussed in section \ref{CavDesign}. We show that such simple and inexpensive rings can be used to tune the zero crossing temperature over a range of about 30 K which is sufficient to eliminate the zero crossing temperature shift of most combined material cavities. In section \ref{CTEMeas} we present CTE measurements verifying the compensation effect of the ULE rings.

\section{Combined material cavities}\label{CombMatCav}

\begin{figure}[t]
\centerline{\includegraphics[width=8.4cm]{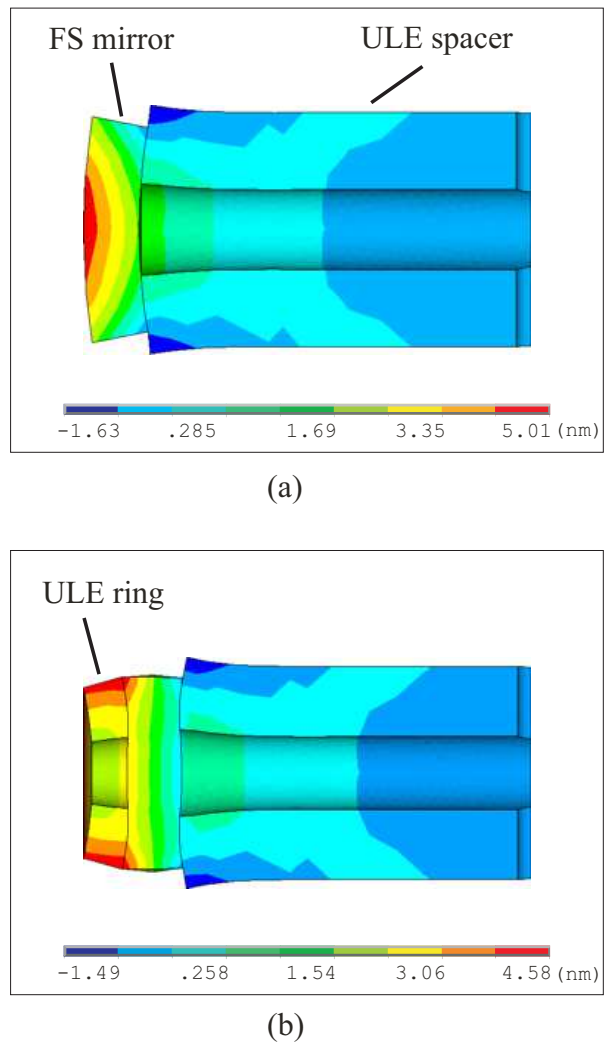}}
\caption{FEM simulations of the elastic cavity deformation after a 1 K temperature step: (a) FS mirror optically contacted to an ULE spacer, (b) additional ULE ring on the back side of the FS mirror to suppress its axial bending. Due to cylindrical symmetry only a quarter of the cavity is simulated. Color scale shows the axial displacement.}\label{FEM-Model} 
\end{figure}

To describe the thermal expansion of combined material cavities we follow an approach discussed in \cite{not95,won97}. We consider a cylindrical cavity consisting of two mirrors with diameter $2 \, R$ optically contacted to a spacer of length $L$. Spacer and mirrors might show different coefficients of thermal expansion $\alpha_\mathrm{s}$ and $\alpha_\mathrm{m}$. Due to the CTE difference a temperature excursion $dT$ results in a difference of the radial expansion between mirror and spacer $dR = (\alpha_\mathrm{m} -\alpha_\mathrm{s}) \, R \, dT$. We assume the optical contact to be perfectly rigid so that the contacted mirror and spacer surfaces can not move with respect to each other. The thermal mirror expansion therefore results in radial mirror stress under which the mirrors bulge in axial direction (Fig.\ref{FEM-Model} (a)). Assuming a linear stress-strain relation, the radial expansion $dR$ and the axial mirror displacement $dB$ are directly connected by a temperature independent coefficient $\delta$ such that $dB = \delta \, dR$. The differential thermal expansion $dL$ of the whole cavity is given by the spacer's thermal expansion $L \, \alpha_\mathrm{s} \, dT$ plus the induced axial displacement of the two mirrors $2 \, dB$. It can be described by an effective cavity CTE with $dL = L \, \alpha_\mathrm{eff} \, dT$ and
\begin{equation}\label{equation_alpha_eff}
\alpha_\mathrm{eff} (T) = \alpha_\mathrm{s}(T) + 2 \delta \, \frac{R}{L} \left[ \alpha_\mathrm{m}(T) - \alpha_\mathrm{s}(T) \right].
\end{equation}
The coupling coefficient $\delta$ depends only on the geometry and the dimensions of the mirrors and the spacer as well as the mechanical properties of mirror and spacer material, but neither on the thermal expansion coefficients nor on the temperature. It can be determined by FEM analysis which has to be done only once for a given mirror spacer geometry and given materials. The effective CTE can then be calculated using Eq.\ref{equation_alpha_eff} for the actual values of $\alpha_\mathrm{m}$ and $\alpha_\mathrm{s}$. 

We have used a commercial FEM package (ANSYS 11.0) to model combined material cavities and simulate their deformations under thermal expansion. Fig.\ref{FEM-Model} (a) shows the FEM result of a cavity composed of an ULE spacer and FS mirrors. It clearly demonstrates the dishing of the FS mirror for a positive temperature step of 1 K. The big mirror deformation is largely supressed by an additional ULE ring as shown in Fig.\ref{FEM-Model} (b). In the simulation the cylindrical spacer is $L=$105.5 mm long and has a diameter of $D=$32 mm. The central bore is 10 mm wide. The FS mirrors are 6.3 mm thick and have a diameter of 25.4 mm. The  Young's modulus $E$ and the Poisson's ratio $\sigma$ are listed in Tab.\ref{tab_mech_properties}.  

\begin{table}[b]
\begin{tabular}{|l|c|c|c|c|}
	\hline
	 & Fused Silica & ULE & Zerodur & ClearCeram \\
	 & (Corning) & (Corning) & (Schott) & (Ohara) \\
	 \hline
	 $E$ (GPa) & 72.7 & 67.6 & 90.3 & 90\\
	 $\sigma$ & 0.16 & 0.17 &  0.243 & 0.25 \\
	\hline
\end{tabular}
\caption{Young's modulus $E$ and Poisson ratio $\sigma$ of low thermal expansion glasses and glass-ceramics taken from the manufacturer data sheets.} \label{tab_mech_properties}
\end{table}

We approximate the instantaneous CTE of ULE glass around its zero crossing temperature $T_\mathrm{0}$ with a quadratic temperature dependence 
\begin{equation}\label{equation_alpha_ULE}
\alpha_\mathrm{ULE}(T) = a \, (T-T_\mathrm{0}) + b \, (T-T_\mathrm{0})^2.
\end{equation}
The linear temperature coefficient $a$ is typically around $1.8 \times 10^{-9}$/K$^2$ and the quadratic coefficient $b$ is as small as $-10 \times 10^{-12}$/K$^3$. For the FEM simulations shown in Fig.\ref{FEM-Model} we have assumed $a = 2.4 \times 10^{-9}/$K$^2$ and $b=0$. The CTE of fused silica shows a value of about $500 \times 10^{-9}$/K at $20\,^{\circ}\mathrm{C}$ and a linear temperature coefficient of $a_\mathrm{FS} = 2.2 \times 10^{-9}$/K$^2$ between $0\,^{\circ}\mathrm{C}$ and $30\,^{\circ}\mathrm{C}$ \cite{ber76}. For all the simulations we approximated the CTE by a constant value of $500 \times 10^{-9}$/K.

To calculate the coupling coefficient $\delta$ we have simulated the cavity deformation for several temperature steps all starting at the zero crossing temperature $T_\mathrm{0}$ and extracted the total cavity length change $\Delta L(T)$ at the mirror center. This length change is described by the effective CTE given in Eq.\ref{equation_alpha_eff} which allows to determine the coupling coefficient by a least square fit. For the cavity model shown in Fig.\ref{FEM-Model} (a) we get $\delta \sim 0.36$.

\begin{figure}[t]
\centerline{\includegraphics[width=8.4cm]{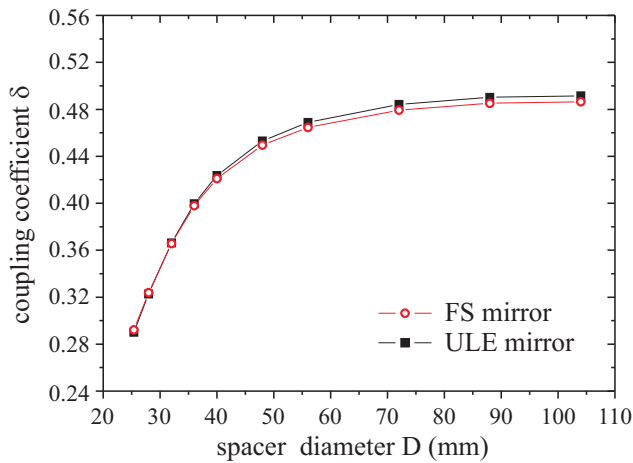}}
\caption{Coupling coefficient $\delta$ versus spacer diameter for a spacer length of 105.5 mm. The mirrors have a diameter of 25.4 mm and a thickness of 6.3 mm.}\label{FEM_coupling_factor} 
\end{figure}

The dependance of the coupling coefficient by the cavity dimensions is illustrated in Fig.\ref{FEM_coupling_factor} for a variation of the spacer diameter. For small spacer diameters the mirror stress is  partially relieved by a deformation of the spacer (Fig.\ref{FEM-Model} (a)). This reduces the longitudinal displacement $dB$ of the mirror center and implies smaller coupling coefficients. For large diameters the coupling coefficient approaches a value of around 0.49.

Such a big coupling factor has a huge impact on the zero crossing temperature of the combined material cavity. Since $\delta$ and the CTE mismatch $\alpha_\mathrm{m}-\alpha_\mathrm{s}$ are both positive, the effective CTE is shifted to higher values in respect to the spacer's CTE (Eq.\ref{equation_alpha_eff}). As the slope of the spacer's CTE $a$ is positive the zero crossing temperature of the cavity is smaller than the zero crossing temperature of the spacer material (Eq.\ref{equation_alpha_ULE}). For the cavity of Fig.\ref{FEM-Model} we get a zero crossing temperature shift of -18 K. 

Even in an all-ULE cavity there can be a CTE mismatch between the spacer and mirror manufactured from different batches. Thus we have  
calculated the coupling coefficient $\delta$ for a cavity completely made of ULE but with different coefficients of thermal expansion of mirror and spacer. As shown in Fig.\ref{FEM_coupling_factor} there is only a small difference in the coupling coefficients for FS and ULE mirrors since both materials show a similar Young's modulus and Poisson's ratio. The Corning specifications for ULE guarantee a mean CTE of $(0 \pm 30) \times 10^{-9}$/K within a temperature range from $5\,^{\circ}\mathrm{C}$ to $35\,^{\circ}\mathrm{C}$. Thus the maximum CTE difference between spacer and mirror is $60 \times 10^{-9}$/K. With the cavity geometry of Fig.\ref{FEM-Model} and the same CTE parameters $a = 2 \times 10^{-9}$/K$^2$ and $b=0$ for both ULE glasses we get a difference between the effective cavity CTE and the spacer's CTE of around $5.5 \times 10^{-9}$/K. This corresponds to a zero crossing temperature shift of about -3 K. 

\section{Temperature insensitive cavity design}\label{CavDesign}

The deformation of the FS mirror and the resulting zero crossing temperature shift can be understood as a consequence of the asymmetry of the mirror-spacer arrangement. The thermal expansion of the FS mirror is only restricted on its front face where it is rigidly connected to the ULE spacer. The free back of the mirror can more easily expand or contract which results in the described mirror dishing. The asymmetry is reduced when additional ULE glass is optically contacted to the mirror's back. For practical reasons, we prefer to use a ring rather than a plate as the smaller area is easier to optically contact and the central hole avoids reflection and distortion of the laser beam at the interface between the FS mirror and the ULE glass \cite{leg08b}. 

\begin{figure}[t]
\centerline{\includegraphics[width=8.4cm]{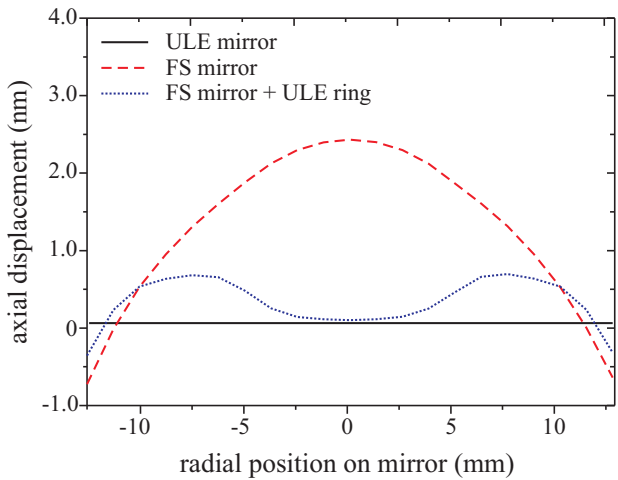}}
\caption{FEM simulation of the axial mirror displacement along a radial line on the mirror surface for a 1 K temperature step starting at $T_0$ of the spacer. An additional ULE ring (thickness 6 mm, inner diameter 9 mm) effectively reduces the mirror displacement around its center (dotted line).} \label{FEM-mirror-displacement}
\end{figure}

Fig.\ref{FEM-Model} (b) shows the FEM simulation of a combined material cavity with an additional ULE ring of the same diameter as the FS mirror and an inner diameter of $d =$ 9 mm. The spacer, the mirror and also the ULE ring are still deformed but the displacement of the mirror center in axial direction is significantly reduced. The mirror deformation becomes more clear in Fig.\ref{FEM-mirror-displacement}. For comparison we also show the axial mirror displacement of a cavity completely made of ULE and a cavity with FS mirrors. With the additional ULE ring the axial mirror displacement at the mirror center is nearly as small as the displacement of the ULE spacer. The corresponding coupling coefficient is reduced from $\delta = 0.366$ without ULE ring to 0.003 for the ULE sandwiched FS mirror. For the cavity model of Fig.\ref{FEM-Model}, the difference between the effective CTE and the CTE of the ULE spacer is therefore decreased from around $43 \times 10^{-9}$/K to about $0.4 \times 10^{-9}$/K. Thus the additional ULE ring reduces the zero crossing temperature shift due to the FS mirror from around -18 K to less than -0.2 K.

\begin{figure}[t]
\centerline{\includegraphics[width=8.4cm]{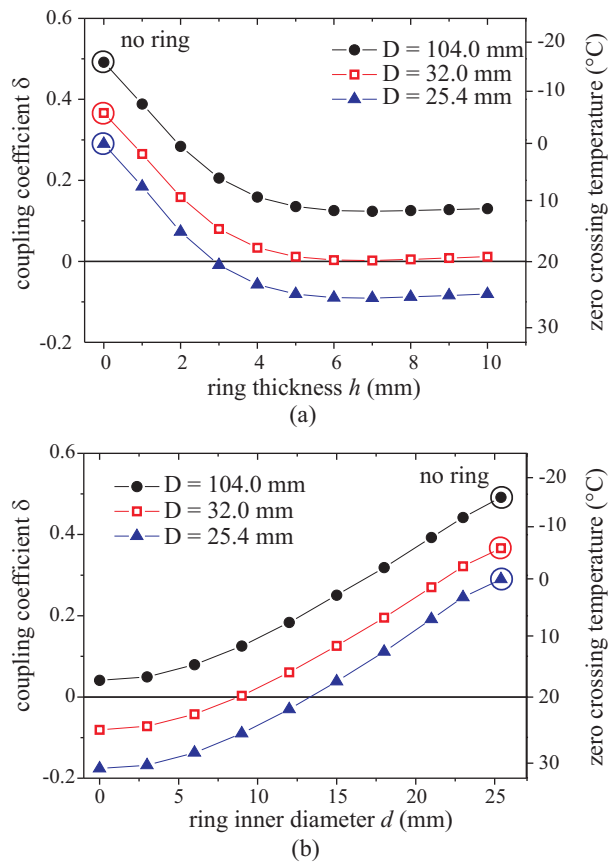}}
\caption{FEM results of the coupling coefficient $\delta$ for an ULE cavity with FS mirrors and additional ULE rings. The coupling coefficient is shown for three different spacer diameters $D$ (25.4 mm, 32 mm and 104 mm). The results at zero ring thickness and inner ring diameter of 25.4 mm, respectively, correspond to no ring at all (circled points). The corresponding zero crossing temperature applies to the cavity mentioned in the text.} \label{FEM-delta-ring}
\end{figure}

The impact of the ULE ring on the coupling coefficient and the cavity's zero crossing temperature depends on the ring dimensions. Fig.\ref{FEM-delta-ring} (a) shows the effect of the ring thickness $h$ for a constant inner ring diameter of $d=$ 9 mm. The coupling coefficient settles at a ring thickness which is comparable to the thickness of the FS mirror. Using thicker rings does not lead to a further reduction of the coupling coefficient. For $\delta = 0$ the effective CTE of the cavity is identical to the CTE of the spacer and there is no zero crossing temperature shift. The big coupling coefficient of the cavity with the largest spacer diameter $D=104.0$ mm can not be completely reduced to zero by the ULE ring considered here. If the spacer diameter $D$ is equal to the diameter of the FS mirror the coupling coefficient can even be negative and the zero crossing temperature of the effective CTE is even higher than the zero crossing temperature of the spacers CTE.
 
Besides the thickness $h$ one can also vary the inner diameter $d$ of the ULE ring. The result shown in Fig.\ref{FEM-delta-ring} (b) for a constant ring thickness of 6 mm illustrates that smaller inner ring diameters will help to decrease the coupling coefficient.

The cavity's zero crossing temperature on the right side of the diagrams is calculated for a 100 mm long ULE spacer with $T_\mathrm{0} = 20.0\,^{\circ}\mathrm{C}$ and FS mirrors of 25.4 mm diameter. We further have assumed a constant FS CTE of $500 \times 10^{-9}$/K and a simplified ULE CTE with $a=2 \times 10^{-9}$/K$^2$ and $b=0$. The temperature scale reveals the tuneability of the zero crossing temperature of the effective CTE. By changing the ring dimensions the zero crossing temperature can be tuned within a temperature range of more than 30 K.  

\section{CTE Measurements}\label{CTEMeas}

We have measured the thermal expansion of four combined material cavities. The four cavities have been made of the same spacer but with different mirror configurations. The cavity dimensions are identical to the ones used for the FEM analysis of section \ref{CombMatCav}. The spacer consists of a low thermal expansion titania silicate glass with material properties similar to Corning's ULE. 

In cavity configuration A the spacer was optically contacted with a pair of ULE mirrors. The pair consists of a plano-concave mirror with a radius of curvature of 1 m and a plane mirror. In configuration B, the plane ULE mirror was replaced by a plane FS mirror of the same dimensions. For the configurations C and D an additional ULE ring with an inner diameter of 12 mm and 9 mm, respectively, was optically contacted to the back side of the FS mirror. The CTE of the ULE rings is not known but should be within the specifications of standard grade ULE (Corning Code 7972).  

During the CTE measurements, the cavity rested on two viton O-rings in a temperature controlled environment inside a vacuum system with a residual pressure of around $10^{-3}$ Pa. The frequency of a He-Ne laser at 633 nm was stabilized to the frequency $\nu_\mathrm{0}$ of a TEM$_\mathrm{00}$ mode of the cavity by a lock-in technique. We monitored the relative frequency change $\Delta \nu / \nu_\mathrm{0}$ in comparison with an iodine stabilized He-Ne laser \cite{rie98a}. The relative length change of the cavity $\Delta L/L$ is equal to $- \Delta \nu / \nu_\mathrm{0}$.  

\begin{figure}[t]
\centerline{\includegraphics[width=8.4cm]{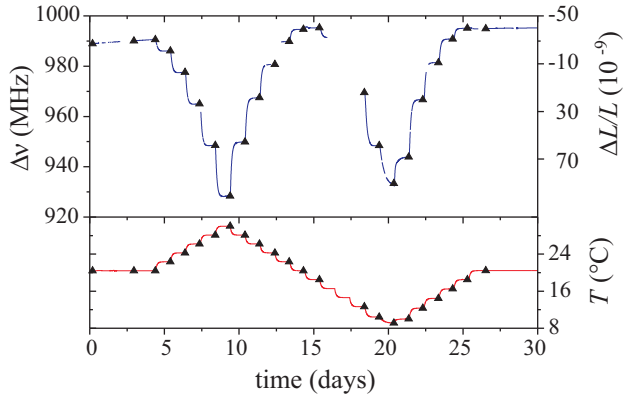}}
\caption{Beat frequency and temperature for the CTE measurement of cavity configuration A over a period of 30 days. The triangles indicate the data used for further analysis.} \label{Fig_CTE_meas_prot}
\end{figure}

\begin{figure}[t]
\centerline{\includegraphics[width=8.4cm]{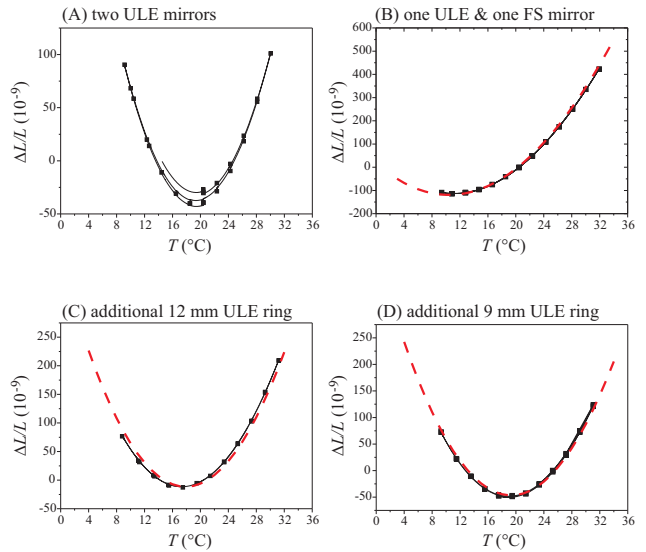}} 
\caption{Thermal relative length change of the four cavity configurations. Shown are the measured values (filled squares) with their fit curves (solid lines) as well as theory curves calculated from the according coupling coefficients $\delta$ (dashed line). }\label{Fig_meas_all}
\end{figure}

The environment temperature of the cavity was changed every 24 hours by temperature steps of 2 K. Cavity temperature and beat frequency were continuously recorded. Fig.\ref{Fig_CTE_meas_prot} shows a typical temperature run and the corresponding beat frequency variation. The cavity temperature follows the 2 K temperature steps with a time constant of about 2.5 h. After 24 h the system was in thermal equilibrium and the relative length change $\Delta L/L$ and the cavity temperature were taken as data points for the thermal expansion shown in Fig.\ref{Fig_meas_all}.  

We use Eq.\ref{equation_alpha_ULE} to describe the effective CTE of the cavities. A linear temporal fractional length drift is taken into account by an additional term $\gamma \, (t-t_0)$. Here $t_0$ is the starting time of the measurement and $\gamma$ is the drift rate. The relative length change of the cavity is then given by 
\begin{equation}\label{equation_Delta_L}
\Delta L/L = a/2 \, (T - T_\mathrm{0})^2 + b/3 \, (T - T_\mathrm{0})^3 + \gamma \, (t - t_\mathrm{0}) + C_\mathrm{0}.
\end{equation}
The CTE parameters of the cavity are obtained by a least square fit of Eq.\ref{equation_Delta_L} to the measured thermal expansion data. The integration constant $C_\mathrm{0}$ depends on the starting conditions and is not relevant to the CTE results. All results are shown in Tab.\ref{tab_CTE_fit_results}. 

The effective CTE of the initial cavity A with two ULE mirrors shows a zero crossing temperature of around $T_\mathrm{0} = 19.4\,^{\circ}\mathrm{C}$. Due to the FS mirror in cavity B the zero crossing temperature of the cavity is shifted to $11.2\,^{\circ}\mathrm{C}$. With two FS mirrors we expect twice the temperature shift and the zero crossing temperature would be around $3\,^{\circ}\mathrm{C}$. With an additional ULE ring of 12 mm inner diameter (cavity C) the temperature shift is partly compensated to  $T_\mathrm{0} = 17.3\,^{\circ}\mathrm{C}$. With a 9 mm ULE ring the cavity shows nearly the same zero crossing temperature as the cavity with two ULE mirrors.

Cavity A shows a relatively large temporal drift of $\gamma = -5.9 \times 10^{-15}$/s, thus $\Delta L/ L$ does not reach its initial value after a full temperature cycle. The temporal drifts of the three remaining cavities are much lower and we attribute the temporal drift of cavity A to the settlement of a bad optical contact of the plane ULE mirror \cite{ber77}. 

Fig.\ref{Fig_meas_all} also shows theory curves calculated from Eq.\ref{equation_alpha_eff} taking into account the two different mirror materials of the cavity. The coupling coefficients $\delta$ are taken from the FEM results shown in Fig.\ref{FEM_coupling_factor} and \ref{FEM-delta-ring}. For a cavity with FS mirrors and a spacer diameter of 32 mm we get $\delta=0.366$. With additional ULE rings of 12 mm and 9 mm inner diameter and 6 mm thickness we get $\delta_\mathrm{12mm}=0.061$ and $\delta_\mathrm{9mm}=0.003$, respectively. We assumed that the spacer's CTE $\alpha_\mathrm{s}(T)$ is equal to the measured effective CTE of cavity A, i.e. the CTE of the ULE mirrors is equal to the one of the spacer. The same CTE was also assumed for the ULE rings of cavity C and D. As CTE differences between ring and mirror act similar to Eq.\ref{equation_alpha_eff}, CTE variations within the specifications of ULE have a minor influence on the compensation effect. The differences between the calculated and the measured zero crossing temperatures are smaller than 0.2 K. This shows that the  model of an effective CTE given by Eq.\ref{equation_alpha_eff} together with FEM simulations of the coupling coefficient $\delta$ is a tool to reliably design the zero crossing temperature of combined material cavities.

\begin{center}
\begin{table*}[t]
{\small
\hfill{}
\begin{tabular}{|l|c|c|c|c|}
\hline
cavity configuration  & A & B & C & D \\
 & 2 ULE & 1 ULE, 1 FS & 12 mm ring & 9 mm ring \\
\hline  
 zero crossing temperature $T_0 \, (^\circ$ C) &  19.38 $\pm$ 0.09 & 11.16 $\pm$ 0.07 & 17.31 $\pm$ 0.03 & 19.0 $\pm$ 0.02 \\
 linear coefficient $a \, (10^{-9}$/K$^2$) & 2.44 $\pm$ 0.02 & 2.7 $\pm$ 0.04 & 2.40 $\pm$ 0.02 & 2.47 $\pm$ 0.01 \\
 quadratic coefficient $b \, (10^{-12}$/K$^3$) & -6 $\pm$ 7 & -15 $\pm$ 2 & -11 $\pm$ 1 & -14 $\pm$ 1 \\
 temporal drift $\gamma \, (10^{-15}$/s) & -5.9 $\pm$ 0.8 & -0.2 $\pm$ 0.15 & -0.6 $\pm$ 0.4 & -1.0 $\pm$ 0.2 \\
\hline
\end{tabular}}
\hfill{}
\caption{CTE parameters of the four cavity configurations. The table shows the results and error bars of a least square fit of Eq.\ref{equation_Delta_L} to the thermal expansion curves.}
\label{tab_CTE_fit_results}
\end{table*}
\end{center}

\section{Conclusion}

The thermal expansion of optical reference cavities made of a low thermal expansion glass spacer and FS mirrors is largely dominated by the deformation of the mirrors. This affects the cavity's CTE and the zero crossing temperature is a few 10 K smaller than the zero crossing temperature of the spacer material. FEM simulations and CTE measurements show that the mirror deformation can be reduced by optically contacting ULE rings on the back surface of the FS mirrors. The ring dimensions can even be used to tune the cavity's zero crossing temperature over a range of more than 30 K. From the good agreement between FEM simulations and experimental results we believe that the compensation technique can also be reliably modelled and successfully applied to more complex cavity geometries, like e.g. tapered cavities \cite{lud07}. 

The influence of the ULE rings to the thermal noise of the cavity was calculated directly from the fluctuation dissipation theorem \cite{Lev98}. The results show that the total thermal noise of a combined material cavity with ULE rings is equal to the cavity's thermal noise without these rings. Details will be given in a forthcoming publication. 

The simple and inexpensive ULE rings greatly simplifies the requirements on thermal shielding and temperature stabilization of combined material reference-cavities with low thermal noise and opens a way to transportable room temperature cavities with a fractional frequency instability down to $10^{-16}$.    

\section*{Acknowledgment}

This work is supported by the European Community's ERA-NET-Plus Programme under Grant Agreement No. 217257, by the ESA and DLR in the project Space Optical Clocks and the Bundesministerium f\"ur Wirtschaft und Technologie within the MNPQ-Transfer program. Thomas Kessler and Uwe Sterr are members of the Centre for Quantum Engineering and Space-Time Research (QUEST). 


\begin{thebibliography}{10}
\newcommand{\enquote}[1]{``#1''}

\bibitem{ros08}
T.~Rosenband, D.~B. Hume, P.~O. Schmidt, C.~W. Chou, A.~Brusch, L.~Lorini,
  W.~H. Oskay, R.~E. Drullinger, T.~M. Fortier, J.~E. Stalnaker, S.~A. Diddams,
  W.~C. Swann, N.~R. Newbury, W.~M. Itano, D.~J. Wineland, and J.~C. Bergquist,
  \enquote{Frequency ratio of {Al$^+$} and {Hg$^+$} single-ion optical clocks;
  metrology at the 17th decimal place,} Science \textbf{319}, 1808--1812
  (2008).

\bibitem{lud08}
A.~D. Ludlow, T.~Zelevinsky, G.~K. Campbell, S.~Blatt, M.~M. Boyd, M.~H.~G.
  de~Miranda, M.~J. Martin, J.~W. Thomsen, S.~M. Foreman, J.~Ye, T.~M. Fortier,
  J.~E. Stalnaker, S.~A. Diddams, Y.~L. Coq, Z.~W. Barber, N.~Poli, N.~D.
  Lemke, K.~M. Beck, and C.~W. Oates, \enquote{Sr lattice clock at $1 \times
  10^{-16}$ fractional uncertainty by remote optical evaluation with a {Ca}
  clock,} Science \textbf{319}, 1805--1808 (2008).

\bibitem{bar05}
A.~Bartels, S.~A. Diddams, C.~W. Oates, G.~Wilpers, J.~C. Bergquist, W.~H.
  Oskay, and L.~Hollberg, \enquote{Femtosecond-laser-based synthesis of
  ultrastable microwave signals from optical frequency references,} Opt. Lett.
  \textbf{30}, 667--669 (2005).

\bibitem{mil09a}
J.~Millo, M.~Abgrall, M.~Lours, E.~English, H.~Jiang, J.~Gu\'ena, A.~Clairon,
  S.~Bize, Y.~L. Coq, G.~Santarelli, and M.~Tobar, \enquote{Ultra-low noise
  microwave generation with fiber-based optical frequency comb and application
  to atomic fountain clock,} Opt. Lett. \textbf{34}, 3707--3709 (2009).

\bibitem{lip09}
B.~Lipphardt, G.~Grosche, U.~Sterr, C.~Tamm, S.~Weyers, and H.~Schnatz,
  \enquote{The stability of an optical clock laser transferred to the
  interrogation oscillator for a cs fountain,} IEEE Trans. Instrum. Meas.
  \textbf{58}, 1258--1262 (2009).

\bibitem{wil08a}
P.~A. Williams, W.~C. Swann, and N.~R. Newbury, \enquote{High-stability
  transfer of an optical frequency over long fiber-optic links,} J. Opt. Soc.
  Am.~B \textbf{25}, 1284--1293 (2008).

\bibitem{jia08}
H.~Jiang, F.~K\'ef\'elian, S.~Crane, O.~Lopez, M.~Lours, J.~Millo,
  D.~Holleville, P.~Lemonde, C.~Chardonnet, A.~Amy-Klein, and G.~Santarelli,
  \enquote{Transfer of an optical frequency over an urban fiber link,} J. Opt.
  Soc. Am.~B \textbf{25}, 2029--2035 (2008).

\bibitem{you99}
B.~C. Young, F.~C. Cruz, W.~M. Itano, and J.~C. Bergquist, \enquote{Visible
  lasers with subhertz linewidths,} Phys. Rev. Lett. \textbf{82}, 3799--3802
  (1999).

\bibitem{sto06}
H.~Stoehr, F.~Mensing, J.~Helmcke, and U.~Sterr, \enquote{Diode laser with
  {1~Hz} linewidth,} Opt. Lett. \textbf{31}, 736--738 (2006).

\bibitem{not06}
M.~Notcutt, L.-S. Ma, A.~D. Ludlow, S.~M. Foreman, J.~Ye, and J.~L. Hall,
  \enquote{Contribution of thermal noise to frequency stability of rigid
  optical cavity via {H}ertz-linewidth lasers,} Phys. Rev.~A \textbf{73},
  031804 (2006).

\bibitem{lud07}
A.~D. Ludlow, X.~Huang, M.~Notcutt, T.~Zanon-Willette, S.~M. Foreman, M.~M.
  Boyd, S.~Blatt, and J.~Ye, \enquote{Compact, thermal-noise-limited optical
  cavity for diode laser stabilization at $1 \times 10^{-15}$,} Opt. Lett.
  \textbf{32}, 641--643 (2007).

\bibitem{web08}
S.~A. Webster, M.~Oxborrow, S.~Pugla, J.~Millo, and P.~Gill,
  \enquote{Thermal-noise-limited optical cavity,} Phys. Rev.~A \textbf{77},
  033847--1--6 (2008).

\bibitem{dub09}
P.~Dub{\'e}, A.~Madej, J.~Bernard, L.~Marmet, and A.~Shiner, \enquote{A narrow
  linewidth and frequency-stable probe laser source for the {$^{88}$Sr$^+$}
  single ion optical frequency standard,} Appl. Phys.~B \textbf{95}, 43--54
  (2009).

\bibitem{num04}
K.~Numata, A.~Kemery, and J.~Camp, \enquote{Thermal-noise limit in the
  frequency stabilization of lasers with rigid cavities,} Phys. Rev. Lett.
  \textbf{93}, 250602--1--4 (2004).

\bibitem{tradename}
Use of tradenames is for informational purpose only.

\bibitem{not05}
M.~Notcutt, L.-S. Ma, J.~Ye, and J.~L. Hall, \enquote{Simple and compact {1-Hz}
  laser system via an improved mounting configuration of a reference cavity,}
  Opt. Lett. \textbf{30}, 1815--1817 (2005).

\bibitem{naz06}
T.~Nazarova, F.~Riehle, and U.~Sterr, \enquote{Vibration-insensitive reference
  cavity for an ultra-narrow-linewidth laser,} Appl. Phys.~B \textbf{83},
  531--536 (2006).

\bibitem{web07}
S.~A. Webster, M.~Oxborrow, and P.~Gill, \enquote{Vibration insensitive optical
  cavity,} Phys. Rev.~A \textbf{75}, 011801(R)--1--4 (2007).

\bibitem{mil09}
J.~Millo, D.~V. Magalh{\~a}es, C.~Mandache, Y.~L. Coq, E.~M.~L. English, P.~G.
  Westergaard, J.~Lodewyck, S.~Bize, P.~Lemonde, and G.~Santarelli,
  \enquote{Ultrastable lasers based on vibration insensitive cavities,} Phys.
  Rev.~A \textbf{79}, 053829 (2009).

\bibitem{ber76}
J.~W. {Berthold III} and S.~F. Jacobs, \enquote{Ultraprecise thermal expansion
  measurements of seven low expansion materials,} Appl. Opt. \textbf{15},
  2344--2347 (1976).

\bibitem{not95}
M.~Notcutt, C.~T. Taylor, A.~G. Mann, and D.~G. Blair, \enquote{Temperature
  compensation for cryogenic cavity stabilized lasers,} J. Phys. D: Appl. Phys.
  \textbf{28}, 1807--1810 (1995).

\bibitem{won97}
E.~K. Wong, M.~Notcutt, C.~T. Taylor, A.~G. Mann, and D.~G. Blair,
  \enquote{Temperature-compensated cryogenic {F}abry - {P}erot cavity,} Appl.
  Opt. \textbf{36}, 8563--8566 (1997).

\bibitem{fox08}
R.~W. Fox, \enquote{{F}abry-{P}erot temperature dependence and surface-mounted
  optical cavities,} Proc. SPIE \textbf{7099}, 70991R (2008).

\bibitem{leg08b}
T.~Legero and U.~Sterr, \enquote{Spiegelbauteil f\"ur einen optischen
  {R}esonator,}  (2008). German patent DE~10~2008~049~367~B3.

\bibitem{rie98a}
F.~Riehle, \enquote{Use of optical frequency standards for measurements of
  dimensional stability,} Meas. Sci. Technol. \textbf{9}, 1042--1048 (1998).

\bibitem{ber77}
J.~W. {Berthold III}, S.~F. Jacobs, and M.~A. Norton, \enquote{Dimensional
  stability of fused silica, invar, and several ultra-low thermal expansion
  material,} Metrologia \textbf{13}, 9--16 (1977).

\bibitem{Lev98}
Y.~Levin, \enquote{Internal thermal noise in the {LIGO} test masses: A direct
  approach,} Phys. Rev.~D \textbf{57}, 659--663 (1998).

\end{thebibliography}

\end{document}